\documentclass[aps,prd,groupedaddress,twocolumn,10pt,nofootinbib]{revtex4-1}

\usepackage{amssymb,amsmath,amsthm}
\usepackage{graphicx}
\usepackage{hyperref}
\hypersetup{colorlinks=false,pdfborder={0 0 0}}
\usepackage{cleveref}

\newcommand{\beq}[1]{\begin{equation}\label{#1}}
\newcommand{\eeq}{\end{equation}}
\newcommand{\bea}[1]{\begin{eqnarray}\label{#1}}
\newcommand{\eea}{\end{eqnarray}}

\newcommand\ylm{Y_{\ell}^{m}}
\newcommand\ylzero{Y_{\ell}^{0}}

\renewcommand{\vec}[1]{\mathbf{#1}}

\renewcommand{\v}[1]{\mathbf{#1}}

\newcommand\veps\varepsilon
\newcommand\vphi\varphi

\newcommand\then\Rightarrow

\begin{document}

\title{The Fortuitous Latitude of the Pierre Auger Observatory and Telescope Array
\texorpdfstring{\\}{}
for Reconstructing the Quadrupole Moment}

\author{Peter B. Denton}
\email{peterbd1@gmail.com}
\author{Thomas J. Weiler}
\email{tom.weiler@vanderbilt.edu}
\affiliation{Vanderbilt University}

\begin{abstract}
Determining anisotropies in the arrival directions of cosmic rays at the highest energy is an important task in astrophysics.
It is common and useful to partition the sky into spherical harmonics as a measure of anisotropy.  
The two lowest nontrivial spherical harmonics, the dipole and the quadrupole, are of particular interest, since these distributions
encapsulate a dominant single source and a plane of sources, as well as offering relatively high statistics.
The best experiments for the detection of ultra high energy cosmic rays currently are all ground-based, 
with highly nonuniform exposures on the sky resulting from the fixed experimental locations on the earth.
This nonuniform exposure increases the complexity and error in inferring anisotropies.
It turns out that there is an optimal latitude for an experiment at which nonuniform exposure does 
not diminish the inference of the quadrupole moment.
We derive the optimal latitude, and find that (presumably by a fortuitous coincidence) this optimal latitude 
runs through the largest cosmic ray experiment, Pierre Auger Observatory (PAO) in the Southern Hemisphere, 
and close to the largest cosmic ray experiment in the Northern Hemisphere, Telescope Array (TA).
Consequently, assuming a quadrupole distribution, PAO and TA can reconstruct the cosmic ray quadrupole distribution to a
high precision, without concern for their partial sky exposure. 
\end{abstract}

\maketitle

\section{Introduction}
\label{sec:introduction}
Detecting any deviation from isotropy in ultra high energy cosmic rays (UHECRs) has been a long sought after goal.
For example, at high energies above the GZK suppression, the anisotropies are expected of reveal the distribution of cosmic ray sources.
One common approach measuring anisotropies is to partition the sky into regions of various sizes with the use of spherical harmonics.
The low order spherical harmonics correspond to simple structures.
The lowest nontrivial order, $\ell=1$, called the dipole, may correspond to a single source where the cosmic rays have been smeared out
by magnetic fields during propagation.
The next order, $\ell=2$, called the quadrupole, can correspond to a planar distribution of sources, such as might be the case if the
sources are spread across the galactic or supergalactic plane.

Several ground based experiments have been detecting UHECRs for several decades.  
Their attempts to measure any sign of anisotropy have yet to find any statistically significant signal.
One fundamental problem with any ground based experiment is that it can only see part of the sky, since the earth is opaque to UHECRs.

Reconstructing the spherical harmonics in general, 
including the quadrupole anisotropy magnitude arising from a planar distribution
of sources, is straightforward with full sky (even nonuniform) coverage.
Partial sky coverage, on the other hand, complicates things dramatically.
While the techniques in the literature for dealing with the dipole on partial sky coverage are fairly simple, those for a quadrupole
are not~\cite{Auger:2012an,Mollerach:2005dv,Billoir:2007kb}.

The largest cosmic ray experiments currently are the Pierre Auger Observatory (PAO) and the Telescope Array (TA).
These experiments will be the focus of this paper.
In section~\ref{sec:spherical} we review spherical harmonics in the present context.
Section~\ref{sec:auger} presents the partial sky exposure function and breaks it down into spherical harmonics; and 
shows that both experiments sit on or near an optimal latitude ($\sim35^\circ$) for inference of Nature's quadrupole strength.
Section~\ref{sec:DiQuadPoles} presents a discussion of how the spherical harmonics mix under partial sky exposure.
In section~\ref{sec:quadrupole} we analytically and numerically show that calculating the strength of the quadrupole anisotropy for
experiments at this optimal latitude does not require considering the effects of partial sky exposure up to a small correction.
We  show the ability to determine if a given quadrupolar distribution is pure or if it has any dipolar contamination in
section~\ref{sec:contamination}.
Section~\ref{sec:dipole} repeats the analytic analysis in the previous section for a dipole, and section~\ref{sec:conclusion} contains
a few conclusions.

\section{Spherical harmonics}
\label{sec:spherical}
The spatial event distribution, $I(\Omega)$, normalized to $\int d\Omega\, I(\Omega)=1$, can be expressed as the sum of spherical
harmonics,
\beq{eq:SphHarm}
I(\Omega)=\sum_{\ell=0}^\infty\sum_{|m|\le\ell}a_\ell^m\ylm(\Omega)\,,
\eeq
where $\Omega$ denotes the solid angle parameterized by the pair of zenith ($\theta$) and azimuthal ($\phi$) angles.
The set $\{\ylm\}$ is complete, and so the expansion Eq.~\ref{eq:SphHarm} is unique.
Besides being complete, the set $\{\ylm\}$ is also orthonormal, with orthonormality condition
\begin{equation}\label{eq:ortho}
\int d\Omega\, Y_{\ell_1}^{m_1}(\Omega)\,Y_{\ell_2}^{m_2*}(\Omega)=\delta_{\ell_1\ell_2}\delta_{m_1m_2}\,.
\end{equation}
The asterisk denotes complex conjugation of the spherical harmonic;
the complex $\ylm$'s satisfy the relation ${\ylm}^*=(-1)^m\,Y_\ell^{-m}$.
The $a_\ell^m$ coefficients contain all the information about the flux distribution.
Inversion of Eq.~\ref{eq:SphHarm} gives the coefficients
\beq{inversion}
a_\ell^m = \int d\Omega\,{\ylm}^*(\Omega)\,I(\Omega)\,.
\eeq
This relation makes it clear that $a_\ell^0$ is real (since $\ylzero$ is real),
and that $a_\ell^m,\ m\ne 0$, is complex (because $\ylm,\ m\ne 0$, is complex).

In practice the observed flux is the sum of Dirac delta functions,
\begin{equation}
\bar I(\Omega)=\frac1N\sum_{i=1}^N\delta(\v u_i,\Omega)\,,
\end{equation}
where $\{\v u_i\}_{i=1}^N$ is the set of $N$ directions of cosmic rays and the Dirac delta function is defined on the sphere in the
usual fashion, $\int f(\v u)\delta(\v u,\v v)d\v u=f(\v v)$ for some function on the sphere $f$.
This allows us to estimate the coefficients in Eq.~\ref{inversion} as
\begin{equation}
\bar a_\ell^m=\frac1N\sum_{i=1}^NY_\ell^{m*}(\v u_i)\,.
\end{equation}

The lowest multipole is the $\ell=0$ monopole term which contains no anisotropy information.
The normalization of the all sky event distribution to unity fixes the value $a_0^0=\frac{1}{\sqrt{4\pi}}$.
Guaranteed by the orthogonality of the $\ylm$'s, the higher multipoles (i.e., $\ell \ge 1$) 
when integrated over the whole sky equate to zero.  
Their coefficients $a_\ell^m$, when nonzero, correspond to anisotropies.

We consider the uncorrected coefficients measured directly by an experiment with a nonuniform exposure function $\omega(\Omega)$. 
They are 
\beq{TildeAmps}
b_\ell^m=\int d\Omega\  Y_\ell^{m*}(\Omega)\,I(\Omega)\, \omega(\Omega)\,,
\eeq
where $\omega(\Omega)$ is the experiment's exposure function normalized such that $\int d\Omega\, \omega(\Omega)=4\pi$.
We also take $\int d\Omega\, I(\Omega)\, \omega(\Omega)=1$.
This estimation and Eq.~\ref{TildeAmps} then fix the observed monopole coefficient to be 
$b_0^0=Y_0^0=\frac{1}{\sqrt{4\pi}}$.
As with the $a_\ell^m$'s, the $b_\ell^0$'s are real and the $b_\ell^m$'s, $m\ne 0$, are complex.
The estimation for the sum of discrete points becomes
\begin{equation}
\bar b_\ell^m=\frac1N\sum_{i=1}^NY_\ell^{m*}(\v u_i)\, \omega(\v u_i)\,.
\end{equation}
We then assume that the continuous coefficients, $a_\ell^m,b_\ell^m$ are well estimated by $\bar a_\ell^m,\bar b_\ell^m$.

The path to clarity here is revealed by also decomposing the exposure function into spherical harmonics, 
\beq{ExposureAmps}
\omega(\Omega)=\sum_{\ell=0}^\infty\,\sum_{|m|\le\ell}c_\ell^m\ylm(\Omega)\,,
\eeq
with coefficients $c_\ell^m$ given by,
\begin{equation}\label{eq:blm}
c_\ell^m=\int d\Omega \, Y_\ell^{m*}(\Omega)\, \omega(\Omega)\,.
\end{equation}
Again, the $c_\ell^0$'s are real and the $c_\ell^m$'s, $m\ne 0$, are complex.
From the normalization of the exposure function given above,
we infer that $c_0^0=4\pi\,Y_0^0 = \sqrt{4\pi}$.
The key ingredient of this paper will be the claim that the $c_2^0$ coefficient, 
the amount of quadrupole in the PAO and TA exposure functions,
is nearly zero (and so can be neglected).
Furthermore, the normalization choice on the directly inferred
event distribution implies that $\sum_{\ell,m} (a^{m*}_\ell c_\ell^m) =1$
where $\sum_{\ell,m}$ is shorthand for $\sum_{\ell=0}^\infty\, \sum_{m=-\ell}^{m=+\ell}$ 
(or in a related notation, for $\sum_{\ell=0}^\infty\, \sum_{|m|\le\ell}$).
Since we have seen that $a_0^0$ and $c_0^0$ are real with a product equal to unity, this constraint may be written as
$\sum_{\ell\ge 1, m} (a^{m*}_\ell c_\ell^m) =0$.

We pause here to collect the inferences of our normalization choices:
The $a_\ell^0,\ b_\ell^0$, and $c_\ell^0$'s are real.
The $a_\ell^m,\ b_\ell^m$, and $c_\ell^m$'s with $m\ne 0$ are complex.
In addition, the monopole coefficients are fixed to be 
$a_0^0 = b_0^0 = \frac{1}{\sqrt{4\pi}}$ and $c_0^0=\sqrt{4\pi}$.
The sum $\sum_{\ell\ge 1, m} (a^{m*}_\ell c_\ell^m)$ is zero.

Inserting Eqs.~\ref{eq:SphHarm} and~\ref{ExposureAmps} into Eq.~\ref{TildeAmps} yields
\beq{TildeAmps2}
b_\ell^m = (-1)^m \,\sum_{\ell_1,m_1} \sum_{\ell_2,m_2} a_{\ell_1}^{m_1}\,c_{\ell_2}^{m_2}
\begin{bmatrix}
\ell_1& \ell_2 & \ell \\
 m_1 & m_2  & -m
\end{bmatrix}\,,
\eeq
where we define our bracket as~\footnote{
The triple $\ylm$~integral is equal to a product of Clebsch-Gordan coefficients, 
\begin{multline*}
\int d\Omega\, Y_{\ell_1}^{m_1}(\Omega)\, Y_{\ell_2}^{m_2}(\Omega)\, Y_{\ell}^{m}(\Omega)=\\
  N(\ell_1,\ell_2,\ell)\times (-1)^{m}\,\langle \ell_1\, \ell_2;\, m_1 m_2\, |\, \ell\, -m \rangle\,,
\end{multline*}
where the normalization factor $N$ depends only on the $\ell$'s and not on the $m$'s:
\begin{equation*}
N=\sqrt{\frac{(2\ell_1+1)(2\ell_2+1)} {4\pi\,(2\ell+1)}}\times \langle \ell_1\,\ell_2;0\,0\,|\, \ell\,0\,\rangle\;.
\end{equation*}
The triple $\ylm$~integral is also related to Wigner's $3j$ symbol.
But our bracket notation is more streamlined for the present problem.
}
\beq{BracketDefn}
\begin{bmatrix}
\ell_1&\ell_2 &\ell_3\\
m_1 & m_2  & m_3
\end{bmatrix}
\equiv
\int d\Omega\, Y_{\ell_1}^{m_1}(\Omega)\, Y_{\ell_2}^{m_2}(\Omega)\, Y_{\ell_3}^{m_3}(\Omega)\,.
\eeq
It is clear from the integral definition of the bracket that the bracket is invariant under the interchange of indices.
As is well-known, this bracket, or triple $\ylm$~integral, 
is non-vanishing only if several important requirements are met~\cite{arfken:math.methods}.
The first is that $m_1+m_2+m_3=0$ (i.e., the $m$-rule).
The next is that $|\ell_i-\ell_j |\le\ell_k\le \ell_i+\ell_j$ for different $i,j,k$ (i.e., the triangle inequality rule).
The third is that $\ell_1+\ell_2+\ell_3$ must be even (i.e., the parity rule). 

Eq.~\ref{TildeAmps2}, relating the inferred anisotropy coefficients ($b_\ell^m$'s) to the 
true coefficients ($a_\ell^m$'s) and exposure coefficients ($c_\ell^m$'s), is completely general.

\subsection{Ground-Based Experiments and Right Ascension}
\label{subsec:RA}
The exposures of ground based experiments are essentially constant in the equatorial coordinate ``right ascension'' (RA).
Therefore, expansions of the exposure have non-vanishing coefficients only 
when the $m$-value of the $c_{\ell_2}^{m_2}$ expansion coefficient is zero, i.e., $m_2=0$.
Thus, for experiments with constant efficiency in RA, we may remove the $m$ summations in Eq.~\ref{TildeAmps2} to get
\beq{TildeAmps3}
b_\ell^m = (-1)^m \,\sum_{\ell_1} \sum_{\ell_2} a_{\ell_1}^m\,c_{\ell_2}^0
\begin{bmatrix}
\ell & \ell_1 & \ell_2 \\
 m  &  -m    & 0      
\end{bmatrix}\,.
\eeq
With $m_2$ identically zero, the $m$-rule then requires $m_1=m$, i.e.
inferred $b_\ell^m$ and true $a_{\ell_1}^m$ will share the same $m$-value.
However, the $\ell$-values will in general differ.

The values of nonzero, independent brackets with $\ell$-values up to four, and one $m$ value equal to zero, 
are listed in Tables~\ref{table:brackets1} and \ref{table:brackets2} of appendix~\ref{appx:brackets}.
It is seen that the brackets remain sizable even as $\ell$ increases.
In particular, the brackets with one $\ell_1$ or $\ell_2$ equal to $\ell$ and the other equal to zero 
retain their value $\frac{1}{\sqrt{4\pi}}$ for all values of $\ell$ by orthonormality.
The cutoff in the summation in Eq.~\ref{TildeAmps3} must therefore come from 
the $a^m_\ell$ and $c_\ell^0$ coefficients.
The set of $a_\ell^m$'s are determined by Nature and are awaiting discovery by our experiments.
The set of $c_\ell^0$'s are determined by the location (latitude) of the ground based experiment,
and by the experiment's opening angle of acceptance on the sky.

Since exposures are relatively smooth functions of declination, fits to exposures will be dominated 
by lower multipoles (small $\ell$ values).  
Moreover, as the ground based experiment moves farther from the equator, the symmetry about the equator of its exposure function
decreases; 
this latter fact diminishes the participation of even parity (even $\ell$) multipoles in the fits.  
Thus, we expect the dominant fitted multipole to be the $\ell=1$ dipole, characterized by $c_1^0$ 
(in tandem with an $\ell=0$ monopole to ensure a positive definite flux across the sky).  
How much the second and higher harmonics 
(the quadrupole $\ell=2$~and sextupole $\ell=3$ are characterized by $c_2^0$ and $c_3^0$) 
contribute is fundamental to the theme of this paper.

\subsection{Particular Cases: Dipole and Quadrupole}
\label{subset:particular}
Particular cases of the general Eq.~\ref{TildeAmps3} are illuminating.
We have seen that the inferred monopole ($\ell=0$) is simply given by normalization to be 
$b_0^0=\frac{1}{\sqrt{4\pi}}$.
The inferred dipole expansion is more interesting.
With $\ell=1$, the triangle rule and the parity rule restrict the values of, say $\ell_2$ relative to $\ell_1$,
to be $\pm 1$.  From Eq.~\ref{TildeAmps3},
the dipole sum becomes
\beq{dipole1}
b_1^m = (-1)^m \sum_{\ell_1} \sum_{Z=\pm 1} a_{\ell_1}^m
	 c_{\ell_1 +Z}^0 \begin{bmatrix}
							 1 & \ell_1 & \ell_1+Z \\
							 m  &  -m    & 0      
							\end{bmatrix} \,.
\eeq
We learn a lesson here, that an inferred dipole can be ``faked'' by a non-dipole multipole when multiplied by a multipole 
component of the exposure differing by one unit of $\ell$.
As with any angular momentum addition, a true multipole and an exposure multipole can add constructively or destructively.

Finally, we write down the expansion for the inferred quadrupole.
Here the triangle rule and the parity rule restrict $\ell_2$ to be equal to $\ell_1$ or to differ from $\ell_1$ by two.
The result is 
\beq{quad1}
b_2^m = (-1)^m\,\sum_{\ell_1} \sum_{Z=0,\pm2} a_{\ell_1}^m\, c_{\ell_1+Z}^0 
							\begin{bmatrix}
							 2 & \ell_1 & \ell_1+Z \\
							 m  &  -m    & 			0      
							\end{bmatrix}\,.
\eeq
Here we learn that an apparent quadrupole can also be faked by a true multipole and a multipole moment of the experimental exposure.
For example, a true monopole event distribution ($a_0^0$) would appear without correction as a quadrupole ($b_2^m$)
if the experimental exposure were quadrupolar ($c_2^0$).
To take a more relevant example, a true dipole distribution observed with a dipole exposure ($c_1^0$) may appear 
as a quadrupole distribution ($b_2^m$).

Experiments with less than $4\pi$~exposure, which includes all ground-based experiments,
are  subject to this ambiguity.  
Assumptions, such as which particular moments Nature chooses to present, must be made.
There are two ways around this ambiguity.  Both ways require all sky coverage.
The first way is to consider space-based experiments, such as the Extreme Universe Space Observatory (EUSO)
proposed for an orbital mission aboard the International Space Station~\cite{Adams:2013vea}.
The second way is to combine data from different experiments so that the whole sky is effectively observed.
The latter method introduces the sticky problem of combining experiments which have different systematic errors.
Combining the data of PAO and TA is an example of the latter approach~\cite{Deligny:2014fxa}.

\section{Partial Sky Exposure Function}
\label{sec:auger}
The exposure functions of ground based experiments cover only part of the sky, 
and are highly nonuniform across that part of the sky that they do see.

\begin{figure}
\centering
\includegraphics[width=\columnwidth]{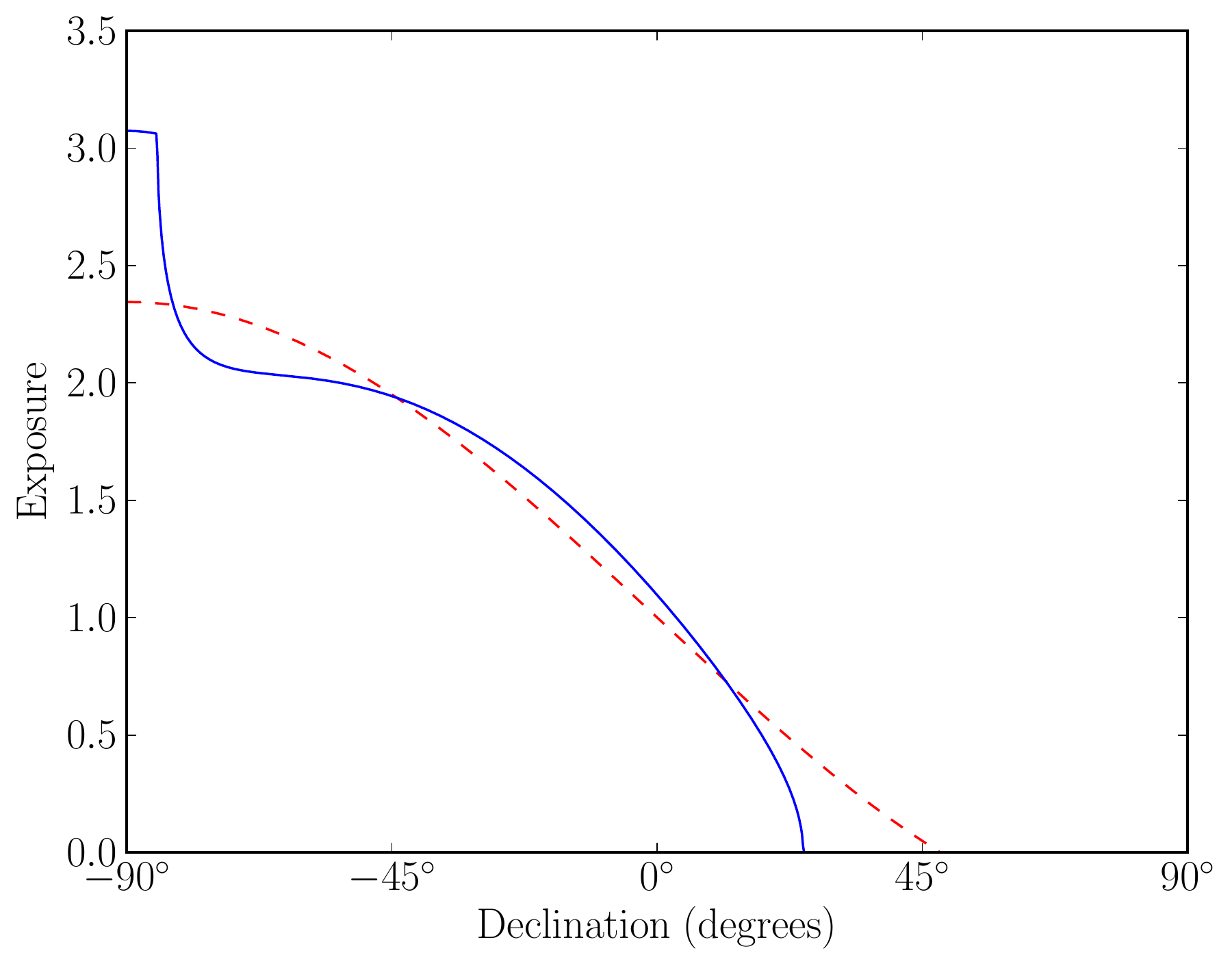}
\caption{The relative exposure of PAO as a function of declination.
Note that the exposure is zero for declinations above $25^\circ$.
The dashed line is the dipole contribution;
the higher order multipole contributions are comparatively small as shown in Fig.~\ref{fig:augers location}.}
\label{fig:auger exposure}
\end{figure}

The relative exposure of a ground based cosmic ray experiment~\cite{Sommers:2000us} is given explicitly by,
\begin{equation}
\begin{gathered}
\omega(\delta)\propto\cos a_0\cos\delta\sin\alpha_m+\alpha_m\sin a_0\sin\delta\\
 \\
{\rm with\ \ } \alpha_m \equiv
\begin{cases}
0&{\rm for}\,\xi>1\\
\pi&{\rm for}\,\xi<-1\\
\cos^{-1}\xi\quad&{\rm otherwise}
\end{cases}\\
 \\
{\rm and\ \ } \xi=\frac{\cos\theta_m-\sin a_0\sin\delta}{\cos a\cos\delta} \,.
\end{gathered}
\label{eq:auger exposure}
\end{equation}
Both PAO and TA are fully efficient at energies above 10 EeV~\cite{Deligny:2014fxa}, well below the threshold typically considered for
UHECRs, $\sim50$ EeV.
The exposure is seen to depend on two experimental parameters, the latitude of the experiment, termed $a_0$ in conventional language,
and the experiment's acceptance angle, $\theta_m$.
Also appearing in the formula is $\delta$, the declination at which the relative exposure is to be calculated.
PAO's latitude and typical acceptance angle are $-35.2^\circ$ and $\theta_m=60^\circ$ 
(the corresponding numbers for TA are $a_0=39.3^\circ$ and the value of $\theta_m$ varies across the
literature, but does not affect this analysis -- we take it to be $\theta_m=45^\circ$ from~\cite{Thomson:2010tc}).
We have assumed that any longitudinal variation in exposure due to weather, down time of the machine, etc., 
is a random process whose average is independent of right ascension (RA).  
Thus, the detector is effectively uniform in RA, and all $c_\ell^m=0$ for $m\neq0$.
The exposure function for a ground based experiment located at PAO's location is shown in Fig.~\ref{fig:auger exposure}.

We turn now to our claim that the quadrupole component of the PAO and TA exposure functions
(i.e., the $c_2^0$ coefficient) is nearly zero in equatorial coordinates at the latitudes of these two experiments.
This claim is neither clear nor automatic. 

To understand the plausibility of vanishing $c_2^0$, notice that if experiments were near the equator, then their exposures would show
a clear $Y_2^0$ shape aligned along the pole.
On the other hand, if the experiments were at a pole, there would be a quadrupole moment in the exposure, although the exposure would
only sense half of it.
Moreover, the value of $c_2^0$ for an experiment at a pole would have the sign of $c_2^0$ opposite to that of 
an experiment at the equator.
Therefore, we infer that there is some latitude $\delta$ in each hemisphere at which $c_2^0$ must vanish.
That unique $|\delta|$ at which $c_2^0$ vanishes turns out to be very near the latitudes of PAO and TA, 
as we now show.

\begin{figure*}
\centering
\includegraphics[width=0.497\textwidth]{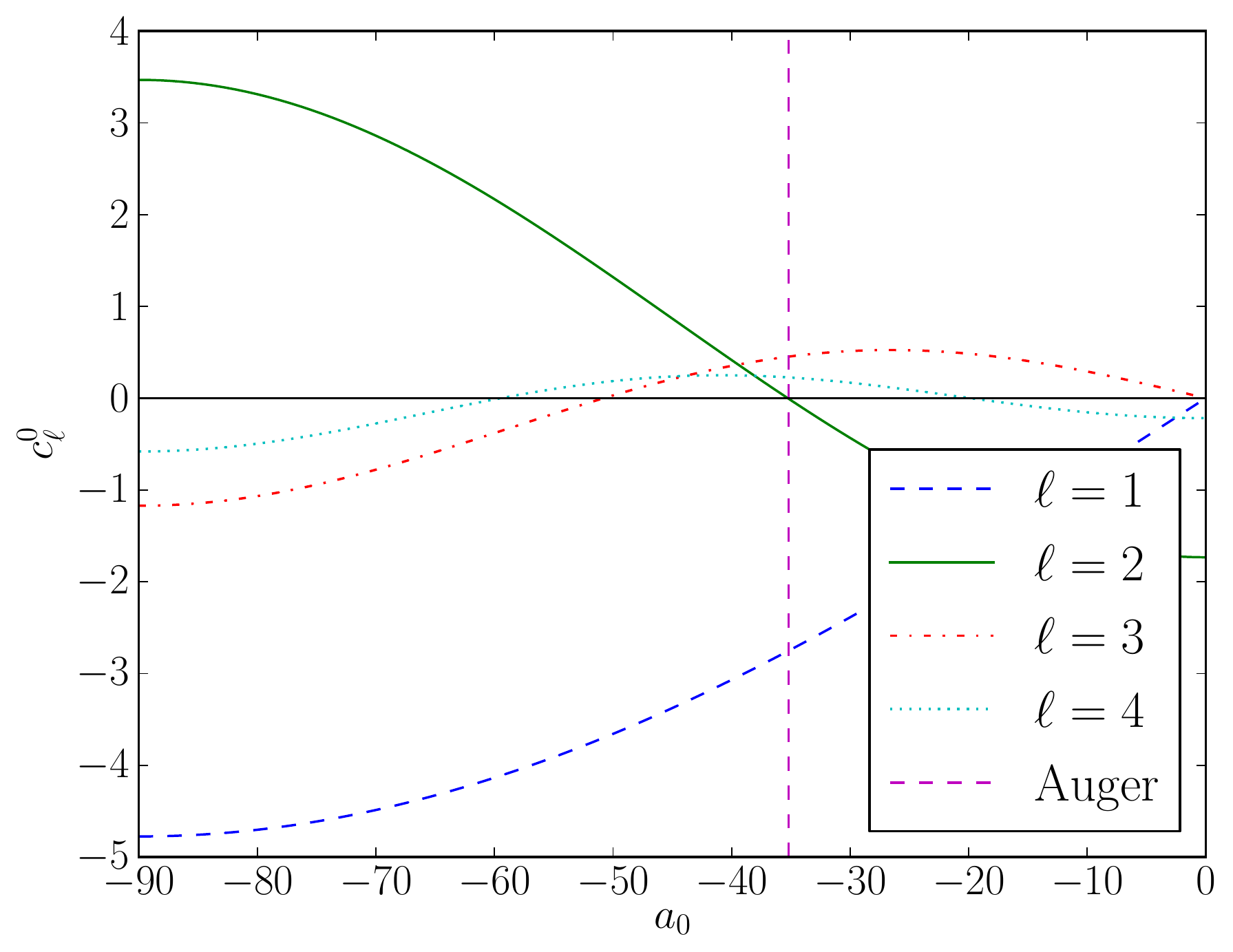}
\includegraphics[width=0.497\textwidth]{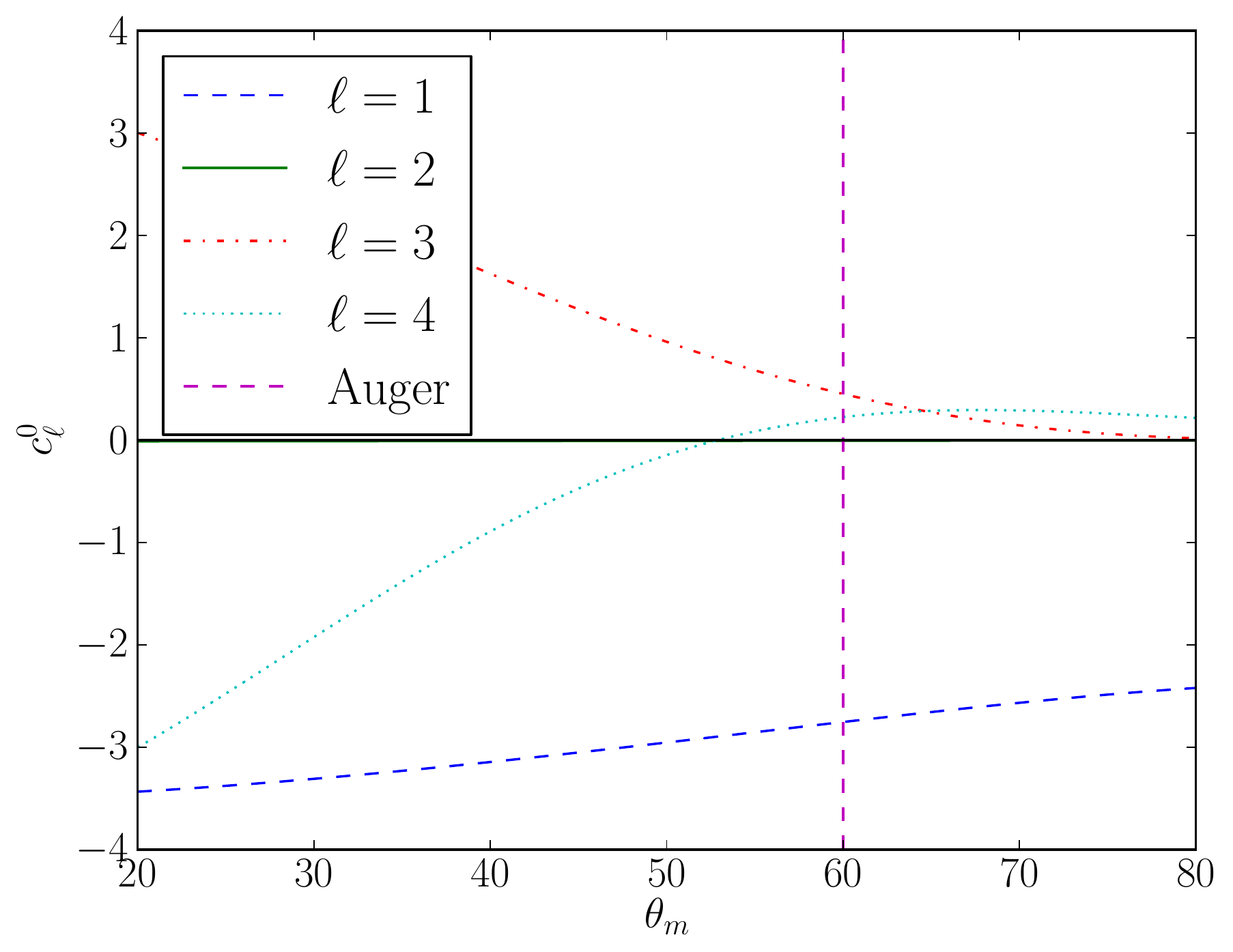}
\caption{In the left panel are the coefficients of the experimental exposure, 
expressed in terms of spherical harmonics at various latitudes, 
with $\theta_m=60^\circ$ fixed.  
In the right panel are the same coefficients in terms of $\theta_m$, 
with $a_0=-35.2^\circ$ fixed to the PAO latitude.
In the right panel, the solid green $\ell=2$ line lies nearly on top of the $c_\ell^0=0$ line,
which implies not only that $c_2^0$ is nearly vanishing at the PAO latitude, 
but also that it is independent of $\theta_m$.}
\label{fig:augers location}
\end{figure*}

In Fig.~\ref{fig:augers location}, we show the first four spherical harmonic coefficients 
for the exposure function of ground based (fixed latitude $\delta$) experiments.
The values of $\ell\neq2$ are included for scale and comparison.
The left panel plots the first four $c_\ell^0$ coefficients versus the experimental latitude $a_0$, in degrees.
It is seen that the $\pm$latitude for which $c_2^0$~vanishes agrees very accurately with the latitudes of PAO and TA.
In fact, PAO's latitude is $-35.21^\circ$ (termed $a_0$ in conventional notation), while the unique latitude at which $c_2^0$ vanishes 
is $-35.26^\circ$.
The difference between these two latitudes is $\sim7$~km -- much smaller than the scale of PAO.
That is, the latitude at which $c_2^0$ vanishes runs right through PAO.

It is also seen that the higher $\ell$ coefficients are small.
The values of the fitted coefficients $c_0^0,\ c_1^0,\ c_2^0, c_3^0, c_4^0$ at the PAO latitude,
and with the PAO opening angle, are 
$\sqrt{4\pi},\ -2.75, \sim 0,\ 0.452,\ 0.226$, respectively.

The dependence on the acceptance angle $\theta_m$ is shown in the right panel of Fig.~\ref{fig:augers location}. 
It is seen that variations in this angle does not affect this vanishing of $c^0_2$ at the PAO latitude.

At PAO's latitude, we see that the largest exposure coefficient in magnitude is $c_1^0$, and that it is negative.
This agrees with the expectation from Fig.~\ref{fig:auger exposure}, which appears to be largely dipolar in nature.
(We also note that at the equator $c_1^0$~vanished, while $c_2^0$ is large as expected.
We comment on this occurrence later in the paper in \S\ref{sec:dipole}.)

A conceptual complication is that the exposure function is necessarily evaluated in equatorial coordinates,
while anisotropies of interest are best considered in galactic coordinates.
Any rotation in coordinates mixes the set of $(2\ell+1)\ \ylm$'s with fixed~$\ell$.
However, it is possible to reconstruct anisotropy information in
the coordinate frame of one's choosing by using the rotationally invariant ``power spectrum,''
defined by the $\ell$--dependent (but not $m$-dependent) set of numbers
$C_\ell \equiv \frac{1}{2\ell+1}\sum_m |a_\ell^m|^2$.
We discuss this power spectrum again in \S\ref{sec:quadrupole}.
In appendix~\ref{appx:power spectrum} we present a proof of the rotational invariance of the $C_\ell$'s. 

\section{Dipoles and Quadrupoles at PAO}
\label{sec:DiQuadPoles}
Armed with the result that the PAO exposure is dominantly dipole ($c_1^0$) in shape,
with a fortuitously small quadrupole ($c_2^0 \ll c_1^0$), and small higher multipoles,
we expand the dipole and quadrupole sums up to order $\ell_2=2$.
This is high enough order to reveal the gift that comes with the justified neglect of $c_2^0$ for PAO and TA.

Applying a cutoff at $\ell_2=2$ in the expansion of the exposure function, Eqs.~\ref{dipole1}, and~\ref{quad1} are
\begin{multline}
b_1^m = \delta_{m0}\,\frac{c_1^0}{4\pi}  +   
			   		a_1^m
			  + (-1)^m \left\{\,a_1^m c_2^0\, \begin{bmatrix}
								 1  &  1   & 2 \\
								 m & -m & 0      
							\end{bmatrix}
\right.\\
\left.
			  + \, a_2^m c_1^0\, \begin{bmatrix}
							 1  &  2  & 1  \\
							 m & -m & 0      
							\end{bmatrix} 
			  + \, a_3^m c_2^0\, \begin{bmatrix}
							 1  &  3  & 2  \\
							 m & -m & 0      
							\end{bmatrix} 
\right\}  \,,
\label{dipole2}
\end{multline}
and
\begin{multline}
b_2^m = \delta_{m0}\,\frac{c_2^0}{4\pi} + a_2^m+
(-1)^m
\left\{
a_1^m c_1^0\,\begin{bmatrix}
			2   &  1  & 1 \\
			m  & -m & 0      
			\end{bmatrix} 
\right. 
\\
+a_2^m c_2^0\,\begin{bmatrix}
			2   &  2  & 2 \\
			m  & -m & 0      
			\end{bmatrix}
+ a_3^m c_1^0\,\begin{bmatrix}
			2   &  3  & 1 \\
			m  & -m & 0      
			\end{bmatrix}
\\
\left.
 + a_4^m c_2^0\,\begin{bmatrix}
			    2 & 4 & 2 \\
			    m & -m & 0      
			   \end{bmatrix}
\right\} \,. 
\label{quad2}
\end{multline}

With $| c_2^0(\delta,\theta_m)|$ set to its minimum value of zero, a very good approximation for the latitudes of PAO and TA,
we get
\begin{multline}
b_1^m =   	\delta_{m0} \frac{c_1^0}{4\pi} +	a_1^m
			  + (-1)^m a_2^m c_1^0\, \begin{bmatrix}
							 1  &  2  & 1  \\
							 m & -m & 0      
							\end{bmatrix} \,,
\label{dipole3}
\end{multline}
and
\begin{multline}
b_2^m = a_2^m + (-1)^m c_1^0
\left\{
a_1^m \,\begin{bmatrix}
			2   &  1  & 1 \\
			m  & -m & 0      
			\end{bmatrix} 
\right. 
\\
\left.
+ a_3^m \,\begin{bmatrix}
			2   &  3  & 1 \\
			m  & -m & 0      
			\end{bmatrix} \right\}\,. 
\label{quad3}
\end{multline}
The dependence of $b_1^m$ on $a_3^m$, 
and the dependence of $b_2^m$ on $a_4^m$, 
has vanished with $c_2^0$ set to zero.
However, the ``raw'' dipole $b_1^m$ still depends on $c_1^0$ and $a_2^m$, in addition to $a_1^m$;
while the ``raw'' quadrupole $b_2^m$ depends on $c_1^0$ and $a_1^m$, and $a_3^m$, in addition to $a_2^m$.

From the fit to the PAO exposure function, the value of $c_1^0$ is $-2.75$.
All of the nonzero brackets in Eqs.~\ref{dipole3} and~\ref{quad3} have a factor of $(-1)^m$, and in absolute value are
$\in[0.18,0.26]$.
Thus the observed $b_\ell^m$, a mixture of the actual $a_\ell^m$'s, actually contain significant influence from
$a_\ell^m$'s with other $\ell$ and $m$ values.

We now consider further simplifications on top of the cutoff in the expansion of the exposure function.
If it is assumed that Nature's distribution of sources is a true dipole (plus monopole, of course), then all $a_\ell^m$ vanish except
for $a_1^m$.
Then the dipole and quadrupole Eqs.~\ref{dipole3} and~\ref{quad3} become simply
\beq{dipole4}
b_1^m = \delta_{m0}\,\frac{c_1^0}{4\pi}  + a_1^m \,,
\eeq
and
\beq{quad4}
b_2^m = (-1)^m a_1^m\,c_1^0\,
				\begin{bmatrix}
				 2   &  1  & 1 \\
				 m  & -m & 0      
				\end{bmatrix} 
 \,,
\eeq
reasserting the notion that the $b_\ell^m$ include a mixture of the $a_\ell^m$.
Here the correction to the dipole term is numerically $-0.219\delta_{m0}$ in Eq.~\ref{dipole4} and in Eq.~\ref{quad4} it is
$b_2^m=-0.694 a_1^m,-0.601 a_1^m,0$ for $|m|=0,1,2$ respectively.
Thus the corrections may well be quite significant.

On the other hand, if it is assumed that Nature's distribution of sources is a true quadrupole (plus monopole), 
then all $a_\ell^m$ vanish except for $a_2^m$, and
the dipole and quadrupole Eqs.~\ref{dipole3}--\ref{quad3} become simply
\beq{dipole5}
b_1^m = \delta_{m0}\,\frac{c_1^0}{4\pi}  + (-1)^m  a_2^m\,c_1^0\, 
					\begin{bmatrix}
					 1  &  2   & 1 \\
					 m & -m & 0      
					\end{bmatrix}	
  \,,
\eeq
and
\beq{quad5}
b_2^m = a_2^m \,.
\eeq
Then the dipole term will be notably nonzero, while the quadrupole will be reconstructed correctly and exactly 
without accounting for the exposure function in any fashion (including the initial cutoff at $\ell_2 > 2$).

The $\bar b_\ell^m$'s are determined experimentally and in the limit of $N\to\infty$, $\bar b_\ell^m=b_\ell^m$.
Thus, Eqs.~\ref{dipole3}--\ref{quad3} could be inverted to yield the desired $a_\ell^m$'s.
That approach -- applying an $\ell_2$ cutoff to the expansion of the exposure function and inverting -- is similar to the $K$-matrix
approach described in~\cite{Billoir:2007kb}.
The mixing of coefficients shows up in the $K$-matrix approach as taking the inverse of a non-diagonal matrix.

The description presented here has two slight advantages over the $K$-matrix approach.
First, the symmetries making some terms equivalent or zero are made explicit by the properties of the bracket object.
Second, the fact that $c_2^0$ is zero or sufficiently small for PAO and TA respectively can be explicitly taken advantage of.
While the $K$-matrix approach is easily extended to arbitrary order in the $\ell_2$ cutoff, the resolution falls off very quickly as the
cutoff $\ell_2$ is increased; 
the practicality of the approach fails for $\ell_2\gtrsim2$.
PAO uses this $K$-matrix approach, but only up to a cutoff of $\ell_2=2$.
PAO obtains a result easily reproduced here.

\quad

In the next section of this paper, we follow a similar path, but we include the full expansion of the exposure function and consider
the case when Nature provides just a quadrupole anisotropy.

\section{Pure Quadrupoles and PAO}
\label{sec:quadrupole}
\begin{figure}
\includegraphics[width=\columnwidth]{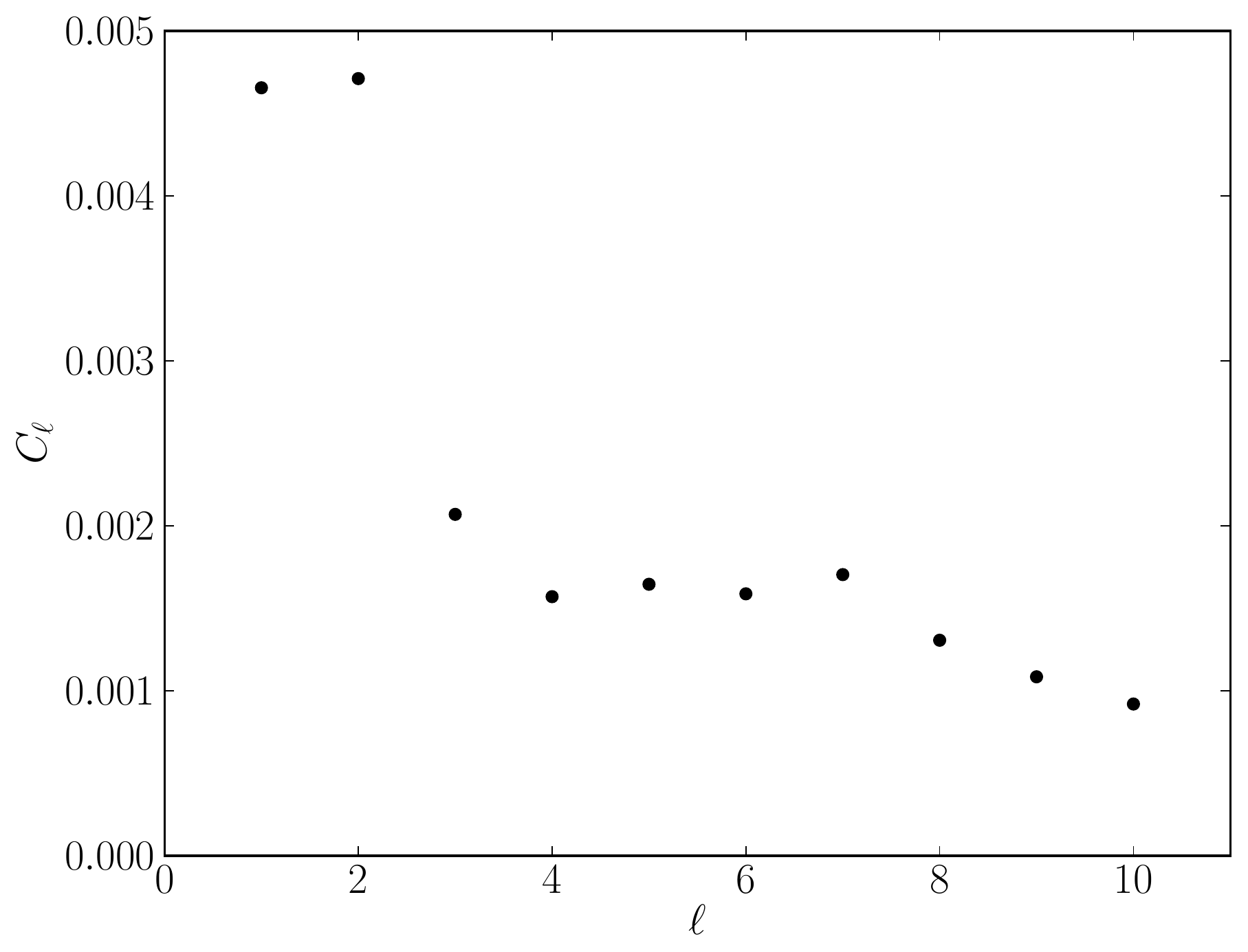}
\caption{The power spectrum (see Eq.~\ref{eq:power spectrum}) for nearby galaxies out to $z=0.028$ (to $d=120$ Mpc) 
weighted according to $1/d^2$.
The 2MRS catalog~\cite{Huchra:2011ii} lists peculiar velocities for the smallest redshifts, and corrects the galactic
latitude $|b|<10^\circ$ cut.
$C_2$ is large because the galaxies roughly form a planar (quadrupolar) structure;
$C_1$ is large because we are not in the center of the super cluster, 
thereby inducing an additional dipole contribution.}
\label{fig:Clgals}
\end{figure}
Quadrupoles are characterized by the $\ell=2$ spherical harmonics.
They are theoretically motivated by planar source distributions, and the presence of magnetic fields.
As such, the quadrupole anisotropy is typically described by just the parabolic $Y_2^0\propto3z^2-1$ spherical harmonic 
oriented along a particular symmetry axis, plus a monopole term to
maintain positive definiteness.
This quadrupole describes a maximum along a band across the sky (at $z\sim$~zero), 
and two minima regions at the poles of the axis of symmetry of the quadrupole
moment ($z\sim\pm 1$).
This quadrupole plus monopole distribution can be written in general as $I(\Omega)=1-C\cos^2\theta$, 
where $\theta$ is the angle measured from the axis of symmetry, and 
$C\in[0,1]$ is some constant. 
The value $C=0$ corresponds to an isotropic monopole, while 
$C=1$ corresponds to the case where the flux goes to zero at the poles of the symmetry axis -- maximum anisotropy.
Such a quadrupole signal would likely be evidence of sources distributed along the galactic or super galactic plane.
The galactic plane is disfavored as a source, since the Larmor radius of protons at energies above the observed dip 
in the cosmic ray spectrum ($10^{18.5}$~eV) should exceed the dimensions of our galaxy.

In either case, galactic or super galactic planes, 
equatorial coordinates are no longer appropriate. 
Galactic or super galactic coordinates are typically used.
These systems are related by rotations, but rotating coordinates has the unfortunate effect of mixing up the 
$m$-values of the spherical harmonics.
A related quantity is the previously mentioned rotationally invariant power spectrum, $C_\ell$. 
The power spectrum of the super galactic plane is shown in Fig.~\ref{fig:Clgals}.
The dipole ($\ell=1$) and quadrupole ($\ell=2$) powers are large, 
as befits a planar distribution viewed from a non-central vantage point.

In the coordinate frame aligned with the quadrupole distribution, 
only $a_2^0$ is nonzero (other than the $a_0^0$ monopole term) for an assumed quadrupolar flux.
Since the power spectrum is defined for particular values of $\ell$, 
in other coordinate systems different values of $\ell$ will not mix, although some of the $a_2^m$'s 
may become nonzero. 
In this regard, the $\ell$ value behaves as that from which it originally came, total angular momentum.

\subsection{Formulas for Pure Quadrupole}
\label{subsec:Formulas Quadrupole}
When a pure quadrupole moment distribution is assumed, 
only the reconstructed $b_{\ell=2}^m$~terms are of interest.
The pure quadrupolar flux distribution is given by 
$I_{\rm Quad}(\Omega)=\frac1{4\pi}+\sum_{m=-2}^2a_2^m\, Y_2^m(\Omega)$.
Then the near vanishing of $c_2^0$ for PAO and TA leads to the simple relation
\begin{equation}
b_2^m=a_2^m(-1)^m\left\{
c_0^0
\begin{bmatrix}
2&2&0\\
m&-m&0
\end{bmatrix}
+
c_4^0
\begin{bmatrix}
2&2&4\\
m&-m&0
\end{bmatrix}
\right\}\,.
\end{equation}
The first bracket is $(-1)^m/\sqrt{4\pi}$ and so the first term is simply $a_2^m$. 
According to values from Table~\ref{table:brackets1} of appendix~\ref{appx:brackets}, 
the second bracket is $f(m)/7\sqrt{4\pi}$, where $f(m)=6,4,1$ for $|m|=0,1,2$, respectively.
Inputing these values, the above expression becomes
\begin{equation}
\label{fortuitous}
b_2^m=a_2^m\left[1+\frac{(-1)^m\, c_4^0\, f(m)}{7\sqrt{4\pi}}\right]\,.
\end{equation}
Using the PAO exposure function yields 
the value of $c_4^0$ fitted to the PAO exposure is 0.226,
so that the final term is $0.0546$, $-0.0364$, and $0.00909$ for $|m|=0,1,2$, respectively.
So $b_2^m\approx a_2^m$ up to a correction of $\lesssim5\%$.
We note that the mixing of the $a_2^m$'s with different $m$~values, mentioned previously, 
actually improves the precision here as the errors are smaller in the
$|m|=1,2$ cases than in the $m=0$ case.

To summarize this section, the fortuitous positioning of PAO and TA at mid-latitudes in the range $\pm$(30 to 40) degrees
presents an exposure with no $c_2^0$~component.
In turn, this allows these experiments to equate the experimental $b_2^0$ with 
the true $a^0_2$ quadrupole coefficient to $\lesssim5\%$, assuming a negligible true dipole contribution,
 without any consideration of the experiment's partial sky exposure.
This conclusion is the fortuitous occurrence referred to in the title of this paper.

The standard technique in the literature to reconstruct the quadrupole moment with full sky exposure is that outlined by Sommers
in~\cite{Sommers:2000us}.
Since PAO and TA's partial sky exposures can be ignored when reconstructing a pure quadrupole, 
we have shown that it is possible to reconstruct the pure quadrupole amplitude using a uniform exposure technique even when
the exposure is nonuniform.
An explicit presentation of a new approach to reconstruct the quadrupole and a comparison with Sommers's approach can be
found in our appendix~\ref{appx:new v Sommers}.
We note that for the success of either approach, 
the experiment must be at or near the optimal latitude (or the experiment must have
uniform full sky exposure for which it is also the case that $c_\ell^0=0$ for all $\ell\ge 1$),
and the true quadrupole must be unaccompanied by other multipoles.

\subsection{Numerical Test of the Pure Quadrupole}
\label{subsec:Numerical Quadrupole}
For numerical confirmation that PAO and TA's exposures do not need to be accounted for when reconstructing the quadrupole anisotropy
measure, we simulate pure quadrupole distributions of cosmic rays and apply a partial sky exposure at various latitudes.
We then reconstruct the pure quadrupole moment.

\begin{figure}
\centering
\includegraphics[width=\columnwidth]{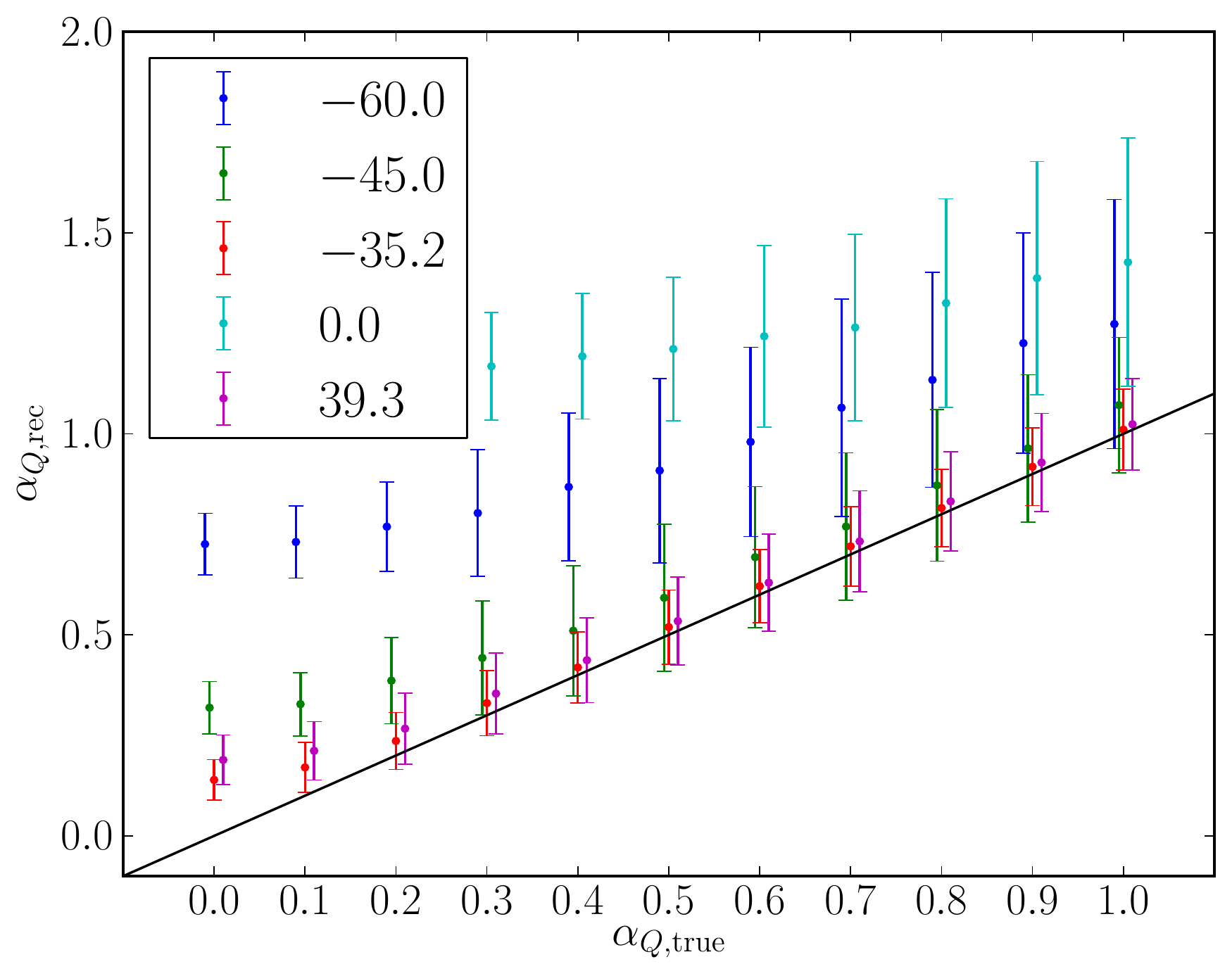}
\caption{Quadrupoles with magnitudes shown on the horizontal axis are injected into an experiment at the shown latitudes.
The error bars for the inferred quadrupoles correspond to one standard deviation over 500 repetitions 
with a different symmetry axis in each repetition.
The black line is $\alpha_{Q,{\rm rec}}=\alpha_{Q,{\rm true}}$.
The behavior of the inference at $a_0=-35.2^\circ$ away from the line at low values of $\alpha_{Q,\rm{true}}$ is due to random fluctuations.
The horizontal shift within one value of $\alpha_{Q,\rm{true}}$ for different latitudes is implemented for clarity only.
}
\label{fig:AQrec_v_AQtrue}
\end{figure}

The process for generating a quadrupole distribution is to first pick a symmetry axis.
Then, we generate cosmic rays with a flux $I(\Omega)\propto1-C\cos^2\theta$ where $\theta$ is the angle between the symmetry axis and
$\Omega$, and $C\in[0,1]$ is some constant.
The standard measure for anisotropy is
\begin{equation}
\alpha=\frac{I_{\max}-I_{\min}}{I_{\max}+I_{\min}}\in[0,1]\,.
\label{eq:alpha}
\end{equation}
For a purely quadrupolar distribution,
\begin{equation}
\alpha_Q=\frac C{2-C}\,.
\label{eq:C to alpha}
\end{equation}
The reconstructed pure quadrupole magnitude is shown in Fig.~\ref{fig:AQrec_v_AQtrue} for several true quadrupole magnitudes and several
latitudes.
In each simulation, a symmetry axis was randomly chosen.
Then 500 cosmic rays were distributed according to the quadrupolar distribution and the experiment's exposure at the given latitude.
This process was repeated 500 times and the mean and standard deviation is shown.
We see that $a_0=-35.2^\circ$ (PAO) reconstructs the quadrupole well.
In addition, $a_0=39.3^\circ$ (TA) reconstructs the quadrupole well (but slightly less so than $a_0=-35.2^\circ$).
In between the two optimal latitudes, $a_0=\pm35.2^\circ$, we see that the ability to simply reconstruct the quadrupole moment
vanishes by the large discrepancy in the $a_0=0^\circ$ reconstruction attempts.
In addition, locating the experiments closer to the poles also nullifies this effect as seen in the $a_0=-60^\circ$ case.
Finally, at $a_0=-45^\circ$ we see how this effect begins to fall off as we move outside the $30^\circ\lesssim|a_0|\lesssim40^\circ$
region.

\section{A Quadrupole Purity Test}
\label{sec:contamination}
In order to determine if a measured quadrupole is pure or if it is tainted by the other spherical harmonics, we propose a
simple statistical test.
We only consider contamination from the $\ell=1$ dipole contribution.
For a given reconstructed $\alpha_Q$ we want to know that the dipole amplitude $\alpha_D$ --our test statistic-- is small.

\begin{figure}
\centering
\includegraphics[width=\columnwidth]{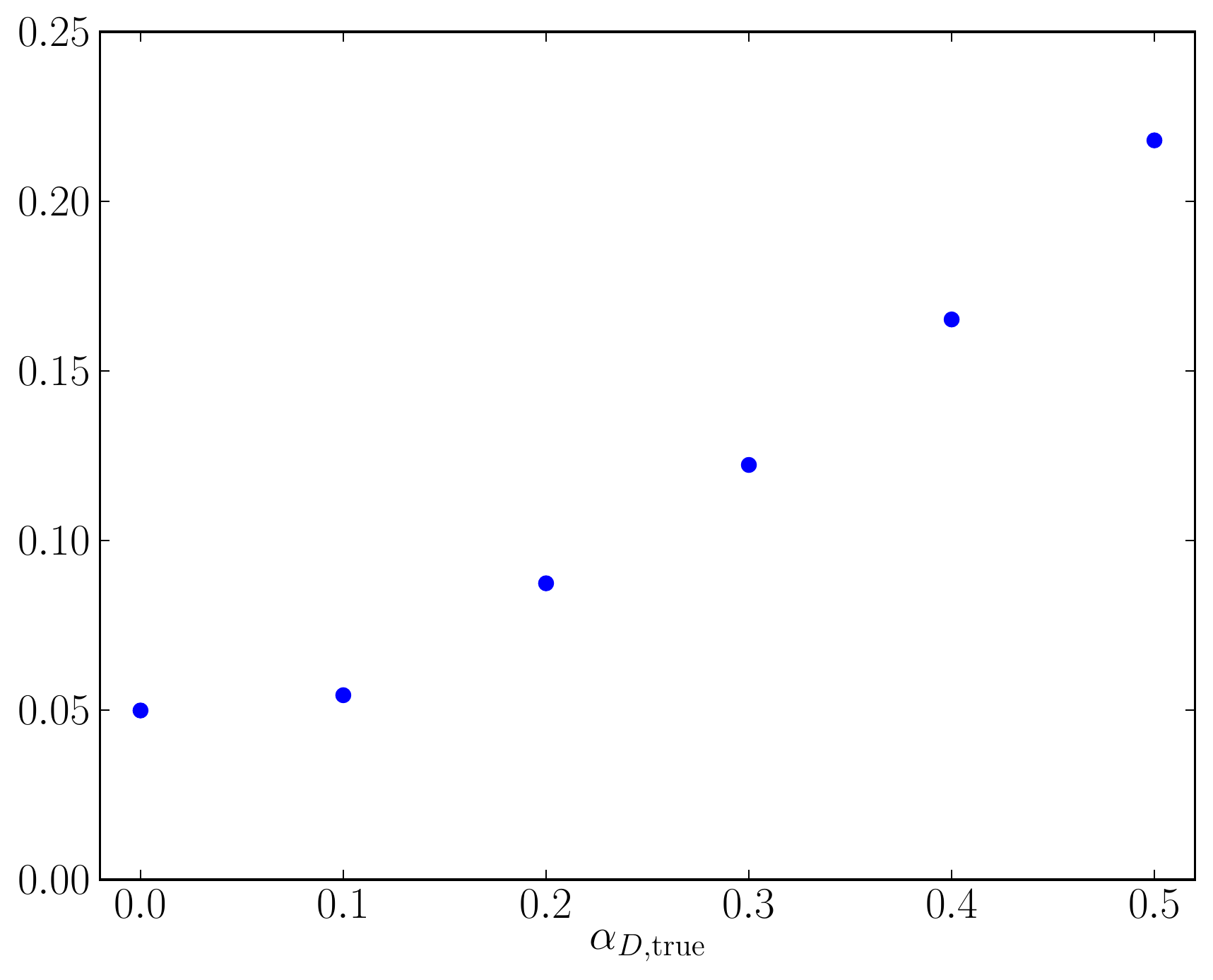}
\caption{
Distributions with 500 cosmic rays, Auger's exposure, maximal quadrupolar anisotropy, and varying dipolar anisotropies were simulated.
$\alpha_D$ was then reconstructed using the $K$-matrix approach.
Plotted on the vertical axis is the fraction of simulations with $\alpha_{D,{\rm rec}}$ not consistent with zero at a 95\% confidence
level.
}
\label{fig:contamination}
\end{figure}

We simulated 500 cosmic rays with Auger's exposure and maximal quadrupole anisotropy, $\alpha_Q=1$.
We then added in increasing amounts of dipole contribution, $\alpha_D$ and reconstructed $\alpha_D$ using the $K$-matrix approach.
After repeating this process 10,000 times we counted how often $\alpha_{D,{\rm rec}}$ was greater than the 95\% limit for the
$\alpha_D=0$ pure quadrupole case.
The results are shown in Fig.~\ref{fig:contamination}.

For example, for a dipole contribution corresponding to $\alpha_D=0.5$ compared with $\alpha_Q=1$, there is a 22\% chance that
$\alpha_D=0$ could be rejected at the 95\% confidence level.
As the contaminating dipole gets weaker, so does this probability.

We see that, as we expected, it is not particularly easy to determine if a quadrupole is pure or not with partial sky coverage.
As shown in Eqs.~\ref{dipole3}--\ref{quad3}, a quadrupolar signal may mimic a dipolar signal and vice versa.

\section{Pure Dipole}
\label{sec:dipole}
For completeness, we address the pure dipole analogy to the pure quadrupole discussion of \S\ref{sec:quadrupole}.
When a pure dipole distribution is assumed, we have 
$I_{\rm Dipole} = \frac{1}{4\pi}+\sum_{m=-1}^{1} a_1^m\,Y_1^m(\Omega)$.
Then the near-vanishing of $c_2^0$ reduces Eq.~\ref{dipole2}  to simply
\beq{dipoleN1}
b_1^m = \delta_{m0}\,\frac{c_1^0}{4\pi}  +   a_1^m\,,
\eeq
the same result as in the $\ell_2$ cutoff case shown in Eq.~\ref{dipole4}.
In the pure quadrupole case, we found that $b_2^m$ was equal to $a_2^m$ times a multiplicative factor
that was within a few percent of unity.
Here we find the equality between $a_1^m$ and $b_1^m$ is exact for $m\ne 0$,
but is broken by an additive factor for $m=0$.
The additive factor is neither large nor small, but rather it is $-0.219$.
This additive factor will also show up with the same relative strength in the power spectrum coefficient $C_1$.

Referring back to Fig.~\ref{fig:augers location}, one sees that an experiment near the equator would have 
vanishing $c_\ell^0$'s except for the quadrupole $c_2^0\sim -1.73$.
It is amusing to ask what pure dipole reconstruction might be possible with such an experimental location.
Eq.~\ref{dipoleN1} is replaced with the following equation:
\begin{align}
b_1^m &= a_1^m \left\{ 1+ (-1)^m\,c_2^0\, 
\begin{bmatrix}
 1  &  1   & 2 \\
 m & -m  & 0      
\end{bmatrix}
\right\} \nonumber \\
 &= a_1^m \left[ 1+ 0.78\,(1-3\,\delta_{m0})\right]\,.
\label{dipoleN2}
\end{align}
Unfortunately, the multiplicative correction from the now nonzero $c_2^0$ coefficient is large.

\section{Conclusion}
\label{sec:conclusion}
We want to stress that the result $b_2^m \approx a_2^m$ as given in Eq.~\ref{fortuitous}
occurs \emph{only} in the special case where the experimental latitude is near $\pm35^\circ$.
At these special latitudes, and only at these special latitudes, is the reconstruction of a pure quadrupole source distribution 
not disadvantaged by partial sky coverage.
By a lucky chance, PAO is located at precisely the correct latitude to benefit from the vanishing of $c_2^0$.
TA's latitude is sufficiently close to $35^\circ$ to also benefit.

If PAO and TA reconstruct both a dipole and a quadrupole simultaneously, 
then the simplification fails, since both experiments have significant dipole moments in their exposure functions.
The ``confusion'' is apparent in the mixing within Eqs.~\ref{dipole3} and \ref{quad3}.
With any source distribution that contains more than a single multipole  (plus monopole, of course), 
the partial sky-coverage inherent in these ground-based experiments exacts a significant price.

Considering that an experiment $5^\circ$ away from the optimal latitude reconstructs the quadrupole well while $10^\circ$ does not, we
claim that experiments at latitudes $30^\circ\lesssim|a_0|\lesssim40^\circ$ can reconstruct a pure quadrupole moment while ignoring
their experiment's particular exposure.

\section{Acknowledgements}
TJW and PBD are supported in part by a Vanderbilt Discovery Grant and a Department of Energy grant, DE-SC0011981.
TJW is also supported by a Simons Foundation Grant, 306329, and acknowledges support from the Munich Institute for 
Astro- and Particle Physics (MIAPP) of the DFG cluster of excellence "Origin and Structure of the Universe,"
and the Aspen Center for Physics, during the preparation of this work.

\appendix

\section{Table of low-multipole brackets}
\label{appx:brackets}
Eq.~\ref{TildeAmps3} describes the general relation between the experimentally-inferred raw multipole coefficients $b_\ell^m$, 
and the true multipole coefficients $a_\ell^m$ and exposure coefficients $c_\ell^0$, in terms of a bracket of the form
\[
\begin{bmatrix}
\ell&\ell_1&\ell_2\\
m&-m&0
\end{bmatrix} \,,
\]
for an experiment with uniform exposure in RA.
In this appendix, we list the nonzero, independent such bracket objects and their values, 
including all $\ell$-values up to four in tables~\ref{table:brackets1} and~\ref{table:brackets2}.
We note that the nonzero brackets remain large in magnitude as $\ell,\ell_1,\ell_2$ are increased.
In particular, they do not generally go to zero with increasing $\ell$-values.

\begin{table}
\centering
\begin{tabular}{||c|c|c|c|c||}
\hline \hline 
 & & & & \\
 $\ell$ & $\ell_1$ & $\ell_2 $  & \ \ \ \ $m$\ \ \ \ \ & $\sqrt{4\pi}\begin{bmatrix} 
\ell & \ell_1 & \ell_2 \\
 m  &  -m    & 0      
\end{bmatrix} $ \\
 \ ($m$)\ \  & \ ($-m$)\ \ & ($m_2=0$) &    &         \\ \hline\hline
\ 0\ \ & \ $\ell_1$\ \ & $\ell_2=\ell_1 $ & 0 & $1$ \\  \hline
  $\ell$ & 0 & $\ell_2=\ell$ & 0 & $1$ \\ \hline
 $\ell$ & $\ell_1=\ell$ & 0 & any $|m| \le \ell$ & $(-1)^m$ \\ \hline\hline
  1 & 1 & 2 & 0               & $\frac2{\sqrt5}$ \\ \hline
      &&		& $\pm 1$ 	  & $\frac1{\sqrt5}$ \\ \hline
1 & 2   & 1 & $\pm 1$   & $-\sqrt{\frac35}$ \\ \hline	
 1 & 2 & 3 & 0  		  & $3\sqrt{\frac3{35}}$ \\ \hline
  	&&		& $\pm 1$   & $\frac3{\sqrt{35}}$ \\ \hline	
 1 & 3 & 2  & $\pm 1$   & $-3\sqrt{\frac2{35}}$ \\ \hline
 1 & 3 & 4  & 0 		  & $\frac4{\sqrt{21}}$ \\ \hline	
  	&&		& $\pm 1$   & $\sqrt{\frac27}$ \\ \hline	
1 & 4 & 3   & $\pm 1$   & $-\sqrt{\frac{10}{21}}$ \\ \hline\hline
 2 & 2 & 2 & 0 				   & $\frac{2\sqrt5}7$   \\ \hline	
	&&		& $\pm 1$  	   & $-\frac{\sqrt5}7$ \\ \hline
	&&		& $\pm 2$		   & $-\frac{2\sqrt5}7$ \\ \hline	
  2 & 2 & 4 &  0 			   & $\frac67$ \\ \hline
  	&&		& $\pm 1$ 		   & $\frac47$ \\ \hline
 	&&		& $\pm 2$  	   & $\frac17$ \\ \hline	
 2 & 3 & 1  & $\pm 1$ 		   & $-2\sqrt{\frac6{35}}$ \\ \hline
 	&&		& $\pm 2$ 	         & $\sqrt{\frac37}$ \\ \hline
 2 & 3 & 3  & 0				   & $\frac4{3\sqrt5}$ \\ \hline
 	&&		& $\pm 1$ 		   & $-\frac13\sqrt{\frac25}$ \\ \hline
 	&&		&  $\pm 2$ 	   & $-\frac23$ \\ \hline
 2 & 4 & 2  & $\pm 1$ 		   & $-\frac{\sqrt{30}}7$ \\ \hline
 	&&		&  $\pm 2$ 	   & $\frac{\sqrt{15}}7$ \\ \hline
 2 & 4 & 4  & 0 			   & $\frac{20\sqrt5}{77}$ \\ \hline
 	&&		& $\pm 1$ 		   & $-\frac{5\sqrt6}{77}$ \\ \hline
 	&&		&  $\pm 2$ 	   & $-\frac{30\sqrt3}{77}$ \\ \hline\hline
\end{tabular}
\caption
{Values for independent, non-vanishing brackets with $\ell=0,1,$~and~2, when $m_2=0$.
When $m=0$ as well, the brackets are symmetric under interchange of any $\ell$ values.
}
\label{table:brackets1}
\end{table}
\begin{table}
\centering
\begin{tabular}{||c|c|c|c|c||}
\hline \hline 
 & & & & \\
 $\ell$ & $\ell_1$ & $\ell_2 $  & \ \ \ \ \ $m$\ \ \ \ \ \ & $\sqrt{4\pi}\begin{bmatrix} 
\ell & \ell_1 & \ell_2 \\
 m  &  -m    & 0      
\end{bmatrix} $ \\
 \ ($m$)\ \  & \ ($-m$)\ \ & ($m_2=0$) &    &         \\ \hline\hline
 3 & 3 & 2 		& $\pm 1$ & $-\frac1{\sqrt5}$ \\ \hline
 	&&			&  $\pm 2$ & $0$ \\ \hline
	&&			& $\pm 3$ & $\frac{\sqrt5}3$ \\ \hline
 3 & 3 & 4 		& 0 & $\frac6{11}$ \\ \hline
 	&&			& $\pm 1$ & $-\frac1{11}$ \\ \hline
 	&&			&  $\pm 2$ & $-\frac7{11}$ \\ \hline
	&&			& $\pm 3$ & $-\frac3{11}$ \\ \hline
3 & 4 & 1  		& $\pm 1$ & $-\sqrt{\frac57}$ \\ \hline
 	&&			&  $\pm 2$ & $\frac2{\sqrt7}$ \\ \hline
	&&			& $\pm 3$ & $-\frac1{\sqrt3}$ \\ \hline
3 & 4 & 3  		& $\pm 1$ & $-\frac{\sqrt{15}}{11}$ \\ \hline
 	&&			&  $\pm 2$ & $-\frac{\sqrt3}{11}$ \\ \hline
	&&			& $\pm 3$ & $\frac{3\sqrt7}{11}$ \\ \hline\hline
4 & 4 & 2  		& $\pm 1$ & $-\frac{17\sqrt5}{77}$ \\ \hline
 	&&			&  $\pm 2$ & $\frac{8\sqrt5}{77}$ \\ \hline
	&&			& $\pm 3$ & $\frac{\sqrt5}{11}$ \\ \hline
	&&			& $\pm 4$ & $-\frac{4\sqrt5}{11}$ \\ \hline
 4 & 4 & 4  	& 0 & $\frac{486}{1,001}$ \\ \hline
 	&& 			& $\pm 1$ & $-\frac{243}{1,001}$ \\ \hline
 	&&			&  $\pm 2$ & $-\frac{27}{91}$ \\ \hline
	&&			& $\pm 3$ & $\frac{81}{143}$ \\ \hline
 	&&			& $\pm 4$ & $\frac{54}{143}$ \\ \hline\hline
\end{tabular}
\caption
{Values for independent, non-vanishing brackets for $\ell=3$~and~4, when $m_2=0$.
When $m=0$ as well, the brackets are symmetric under interchange of any $\ell$ values.
}
\label{table:brackets2}
\end{table}

\section{Proof of the rotational invariance of the power spectrum}
\label{appx:power spectrum}
While calculating the coefficients of the spherical harmonics, the $a_\ell^m$'s, is useful, 
the $a_\ell^m$'s suffer the disadvantage that they are frame dependent.
The spherical harmonics coefficients, the $a_\ell^m$'s given in Eq.~\ref{eq:SphHarm}, are clearly coordinate dependent,
as a simple rotation in the $\phi$ coordinate will change the $e^{im\phi}$ part of the spherical harmonic for $m\neq0$, 
and a rotation in the $\theta$ coordinate will change the associated Legendre polynomial part $P_\ell^m(\theta)$ for $\ell\neq0$.
So only the $\ell=m=0$ monopole coefficient is coordinate independent.

To combat this problem of rotational non-invariance, 
the power spectrum, defined by
\begin{equation}
C_\ell\equiv\frac1{2\ell+1}\sum_{m=-\ell}^\ell|a_\ell^m|^2\,,
\label{eq:power spectrum}
\end{equation}
is often invoked.
While it may be intuitive that the $C_\ell$ should be rotationally (coordinate) invariant, this attribute is not obvious.
The purpose of this appendix is to prove the rotational invariance of the power spectrum.

For a discrete set of sources, the normalized intensity function is $I(\Omega)=\frac{1}{N}\sum_{i=1}^N\delta(\vec{u}_i,\Omega)$.
In terms of spherical harmonics, one finds that the spherical harmonic coefficients are given by
\beq{discrete alm}
\bar a_\ell^m=\frac1N\sum_{i=1}^NY_\ell^{m*}(\vec{u}_i)\,,
\eeq
where $\vec{u}_i$ is the unit direction vector to the $i$th cosmic ray, $1\le i \le N$.

To construct the estimation of the associated power spectrum, we square these $a_\ell^m$'s followed by a sum over $m$:
\begin{align}
\bar C_\ell&\equiv\frac{1}{2\ell+1} \sum_{|m|\le\ell} |\bar a_\ell^m |^2 \nonumber \\
&=\frac{1}{N^2(2\ell+1)}\sum_{|m|\le\ell} \left| \sum_{i=1}^N Y_\ell^{m*}(\vec{u}_i)\right|^2\,.
\end{align}
Since the sums are finite they may be rearranged and expanded to
\begin{align}
\bar C_\ell=&\frac1{N^2(2\ell+1)}\sum_{i=1}^N\sum_{|m|\le\ell}|\ylm(\vec{u}_i)|^2+ \nonumber \\
&+\frac2{N^2(2\ell+1)}\sum_{i<j}\sum_{|m|\le\ell}Y_\ell^{m*}(\vec{u}_i)\ylm(\vec{u}_j)\,.
\label{eq:cl ylm big sum}
\end{align}
The addition formula~\cite{arfken:math.methods} for spherical harmonics is
\begin{equation}
P_\ell(\v x\cdot\v y)=\frac{4\pi}{2\ell+1}\sum_{|m|\le\ell}Y_\ell^{m*}(\v x)\ylm(\v y)\,,
\label{eq:m sum different}
\end{equation}
where $P_\ell(\cos\theta)$ is the Legendre polynomial.
Since $P_\ell(1)=1$, setting the unit direction vectors $\v x$ and $\v y$ to be equal in Eq.~\ref{eq:m sum different}, 
one gets
\begin{equation}
\frac{2\ell+1}{4\pi}=\sum_{|m|\le\ell}|\ylm(\v x)|^2\,.
\label{eq:m sum same}
\end{equation}
Combining Eqs.~\ref{eq:cl ylm big sum},~\ref{eq:m sum different}, and~\ref{eq:m sum same} gives
\begin{equation}
\bar C_\ell=\frac1{4\pi N}+\frac1{2\pi N^2}\sum_{i<j}P_\ell(\v u_i\cdot\v u_j)\,.
\end{equation}
Experimentally only $\bar a_\ell^m$ and $\bar C_\ell$ may be measured, but these are estimates of their continuous counterparts
$a_\ell^m,C_\ell$ respectively.
Finally, since inner products are invariant under rotations, the $C_\ell$ are also invariant under rotations.
This completes the proof.

\section{Our approach to calculating the quadrupole moment}
\label{appx:new v Sommers}
Sommers outlines one method for calculating the anisotropy from a pure quadrupolar distribution~\cite{Sommers:2000us}.
His approach assumes full sky (possibly nonuniform) exposure.
Based on the results from this paper, that same approach can be applied for PAO and TA by ignoring their exposure.

We present here the explicit description of a new approach, based on the results of this paper, that is similar to the $K$-matrix
approach~\cite{Billoir:2007kb}, but explicitly takes advantage of the fact that $c_2^0=0$ and the fact that the $a_\ell^m$ don't mix
as shown in section~\ref{sec:quadrupole} for a purely quadrupolar distribution.

Since the power spectrum is rotationally invariant as shown in appendix~\ref{appx:power spectrum}, we can consider a coordinate frame
that aligns the $z$-axis with the symmetry axis of the quadrupole.
Then only $a_0^0$ and $a_2^0$ are nonzero.

\begin{figure}
\includegraphics[width=\columnwidth]{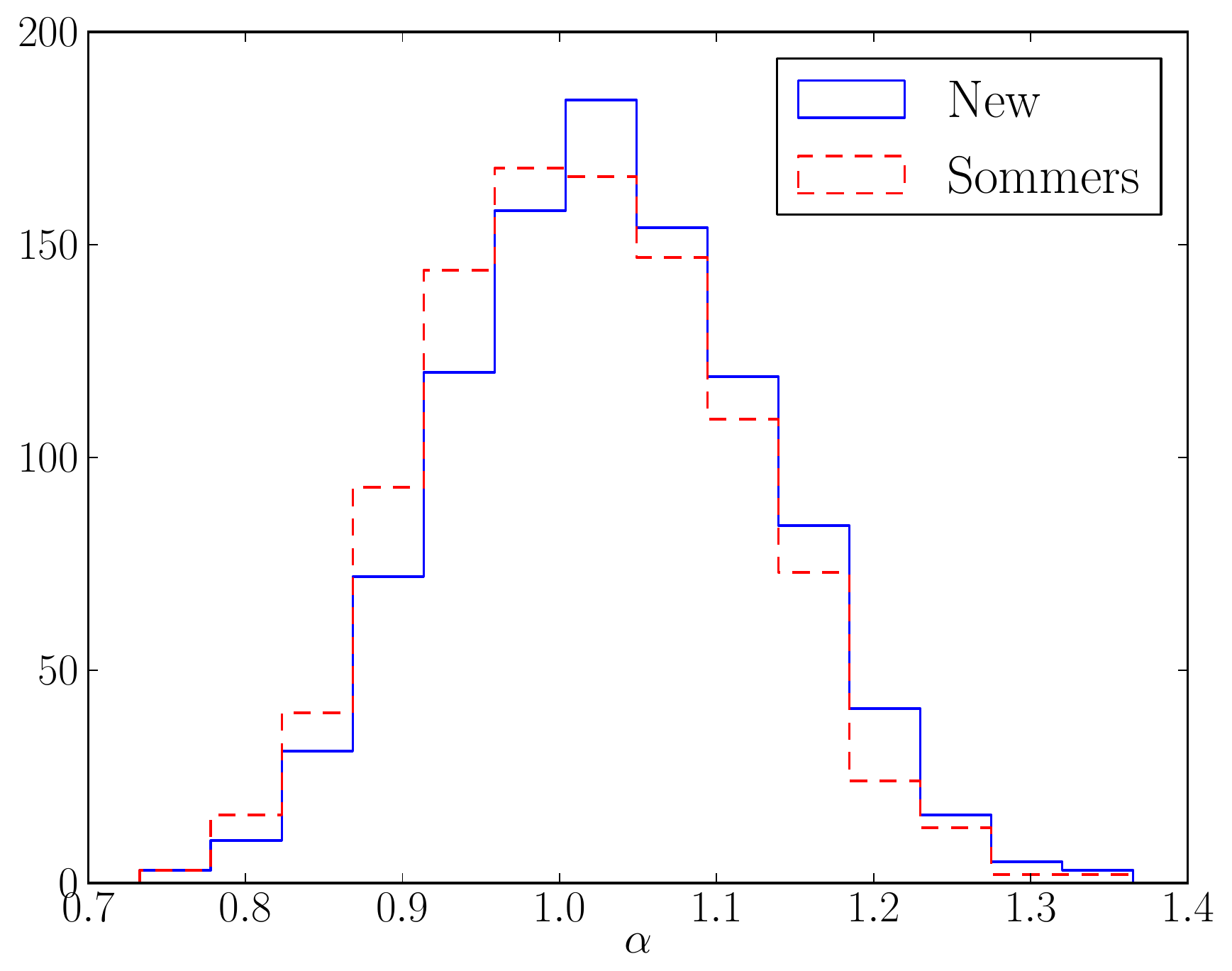}
\caption{500 directions were simulated with a quadrupolar distribution aligned in a random direction, and with PAO's
exposure.
Here, $\alpha_Q$ is set to unity.
The quadrupole strength was then reconstructed with each of the two techniques.
This process was repeated 1000 times.
Both approaches are correlated - that is when one approach gives a small value of $\alpha$, the other will as well.}
\label{fig:new v Sommers}
\end{figure}

In terms of Cartesian coordinates,
\begin{equation}
Y_2^0=A(3z^2-1)\,,
\label{eq:y20}
\end{equation}
with $A=\frac14\sqrt{\frac5\pi}$.
The intensity function in Cartesian coordinates aligned with the quadrupole axis is
\begin{equation}
I=B(1-Cz^2)\,,
\end{equation}
where the normalization requirement sets $\frac1B=4\pi(1-\frac C3)$.
We then invert Eq.~\ref{eq:y20} to write the intensity as a function of $Y_0^0$ and $Y_2^0$.
\begin{equation}\label{eq:I}
I=\frac1{\sqrt{4\pi}}Y_0^0-\frac{BC}{3A}Y_2^0\,.
\end{equation}
In the coordinate frame where the quadrupole symmetry axis is aligned with the $z$ axis, only $a_0^0$ and $a_2^0$ are nonzero and are
given by Eq.~\ref{eq:I}.
Then $a_2^0=-\sqrt{5C_2}$ where we used the definition of the power spectrum from Eq.~\ref{eq:power spectrum} and the sign is because
$a_2^0$ in Eq.~\ref{eq:I} is negative definite.

Then we have a prescription to find $\alpha_Q$ as defined in Eq.~\ref{eq:alpha}.
First calculate the power spectrum for $\ell=2$ in any coordinate frame for data with either full sky coverage or from experiments at
latitudes $30^\circ\lesssim|a_0|\lesssim40^\circ$.
Then get $a_2^0$ and then find $C$ as described in Eq.~\ref{eq:I} and this gives $\alpha_Q$ from Eq.~\ref{eq:C to alpha}.

We compare this approach to that described by Sommers in~\cite{Sommers:2000us}.
We note that this new technique provides no directional information which can be calculated from the eigenvectors of the $Q$ tensor as
described by Sommers.
A plot of the comparison is shown in Fig.~\ref{fig:new v Sommers} of the reconstructed quadrupole amplitude.
The results are consistent with each other.
In particular, the means and standard deviation for each approach are $1.033\pm0.099$ for this new approach and $1.017\pm0.098$ for
Sommers's approach, compared to a correct value of $1$.
The non-negligible deviations in the reconstructed values are not due to the small corrections mentioned in Eq.~25, rather they are
due to the fact that a discrete sampling with finite $N$ number of delta functions of a continuous distribution tends to lend itself to
larger anisotropies.

\bibliography{Quad_Auger}

\end{document}